\documentstyle[aps,prl,preprint,psfig,floats]{revtex}
\tightenlines
\begin{document}
\draft
\preprint{UNDQCG-02-02}
\title{Imperfect Detectors in Linear Optical Quantum Computers}
\author{Scott Glancy, J. M. LoSecco, H. M. Vasconcelos and C. E. Tanner}
\address{Department of Physics, University of Notre Dame, Notre Dame, Indiana 46556}
\date{\today}
\maketitle
\begin{abstract}
We discuss the effects of imperfect photon detectors suffering from loss and noise on the reliability of linear optical quantum computers.  We show that for a given detector efficiency, there is a maximum achievable success probability, and that increasing the number of ancillary photons and detectors used for one controlled sign flip gate beyond a critical point will decrease the probability that the computer will function correctly.  We have also performed simulations of some small logic gates and estimate the efficiency and noise levels required for the linear optical quantum computer to function properly.\\
\end{abstract}


\narrowtext
Theoretically quantum computers can perform some types of calculations much faster than classical computers \cite{DeutschFeynman}, but the technological difficulties of manipulating quantum information have so far prevented researchers from constructing a quantum computer which is able to perform useful tasks.  One of the first proposals for the construction of a quantum computer envisioned a single photon traveling through a network of beam splitters and phase shifters.  Unfortunately for this single photon computer to encode and manipulate $n$ qubits requires on the order of $2^n$ optical modes, beam splitters, phase shifter, and photon detectors. \cite{Reck} The durability of the optical qubits and the simplicity of the linear optical components make this an attractive design for building small (a few qubits) quantum computers, but the exponential growth in the number of components makes this sort large scale quantum computer economically impractical.

This problem of exponential growth seems to have been eliminated by Knill, Laflamme and Milburn in \cite{KLM1,KLM2,KLM3}.  In their proposal (outlined below) a single qubit is represented by the presence of a photon in one of two optical modes.  Two qubits interact by passing through a network of beam splitters and phase shifters along with some ancillary photons. The result of this interference will be a very complicated entangled state, but after the number of photons in the ancillary modes is measured the state will collapse. Depending on the result of the measurement, this process may have performed the desired interaction (the controlled sign flip) or some undesired interaction, in which case the computation has failed.  Knill, Laflamme, and Milburn also show that the probability that the measurement process reports that the operation has succeeded may be increased arbitrarily close to one by increasing the number of ancillary optical modes, photons, beam splitters, etc. This proposal keeps many of the advantage of earlier linear optical quantum computers while eliminating the exponential growth of the number of optical components with the number of qubits. The disadvantage of this scheme is that its calculations succeed only probabilistically, and the cost of improving the probability of success is the use of large (but subexponential) numbers of ancillary photons and photon detectors.

The goal of this paper is to investigate the performance of linear optical quantum computers (LOQC) whose photon detectors have inefficiency and noise. First, we will review the LOQC scheme proposed by Knill, Laflamme, and Milburn.  We then discuss a computer simulation of LOQC with imperfect detectors.  We also examine near deterministic quantum teleportation with inefficient detectors.

\section{Ideal Linear Optical Quantum Computing}
The goal of LOQC is to build a quantum computer that uses photons to encode information, but does not require any nonlinear medium to allow the photons to interact. Instead, we will only use single photon sources, beam splitters, phase shifters, and photon detectors which are able to distinguish between $0$, $1$, $2$, ... photons. According to the original scheme for LOQC, each qubit is represented by a single photon in a superposition of two optical modes.  The conventional notation is
\begin{equation}
|Q\rangle=a|0_1,1_2\rangle+b|1_1,0_2\rangle.
\end{equation}
Here the logical value of $0$ corresponds to the eigenstate that has zero photons in optical mode $1$ and one photon in optical mode $2$, and the logical $1$ state has one photon in mode $1$ and zero photons in mode $2$. We will describe a phase shifter using the phase angle $\phi$ that it imparts to a single photon, and $\hat{{\bf P}}_{\phi}$ is the unitary transformation that acts on the state of the photons traveling through the phase shifter ${\bf P}_{\phi}$. For example $\hat{{\bf P}}_{\phi}|0\rangle=|0\rangle$, and $\hat{{\bf P}}_{\phi}|1\rangle=e^{i\phi}|1\rangle$. Beam splitters ${\bf B}_{\theta,\phi}$ are similarly described using two angles, $\theta$ and $\phi$. They are usually represented by the rotation matrix that transforms the photon creation operators $a^{\dagger}_n$ into their primed versions in the Heisenberg picture
\begin{equation}
\left(
\begin{array}{c}
{a^{\dagger}_1}^{\prime} \\
{a^{\dagger}_2}^{\prime}
\end{array}
\right)
=
\left(
\begin{array}{cc}
\cos(\theta) & e^{i\phi}\sin(\theta) \\
-e^{-i\phi}\sin(\theta) & \cos(\theta)
\end{array}
\right)
\left(
\begin{array}{c}
{a^{\dagger}_1} \\
{a^{\dagger}_2}
\end{array}
\right)
\end{equation}
Because any large unitary matrix can be written as a direct product of two by two unitary matrices, we can describe the action of many beam splitters on any number of optical modes by giving the matrix that is the direct product of the matrices describing each beam splitter and phase shifter. Suppose then that $G$ is such a matrix describing the transformation of many optical modes; the elements of $G$ are written as $g_{ij}$; and $\hat{G}$ is the unitary operator acting on photon states in the Schr\"{o}dinger picture. To find how a Fock state is transformed by $\hat{G}$ we must compute
\begin{equation}
\hat{G}|n_1,n_2,...,n_m\rangle=\left(\prod^{m}_{i=1}\frac{1}{\sqrt{n_i!}}\left(g_{i1}a^{\dagger}_1+g_{i2}a^{\dagger}_2+...+g_{im}a^{\dagger}_m\right)^{n_i}\right)|0_1,0_2,...,0_m\rangle
\end{equation}
This formalism allows us to generalize the concept of beam splitters to include a much wider class of devices, all of which can be constructed by collections of single mode phase shifters and conventional two mode beam splitters \cite{Leonhardt}.

The first element of a LOQC is a device called $NS_{-1}$ that will shift the phase of a mode containing two photons, but will leave a single photon undisturbed. $NS_{-1}$ performs the transformation
\begin{eqnarray}
|0_0\rangle & \rightarrow & |0_0\rangle \\
|1_0\rangle & \rightarrow & |1_0\rangle \\
|2_0\rangle & \rightarrow & -|2_0\rangle.
\end{eqnarray}
Performing this operation requires the preparation of two ancillary modes containing the state $|1_1,0_2\rangle$. The three modes are then sent through a network of beam splitters described by the matrix
\begin{equation}
\left(
\begin{array}{ccc}
1-\sqrt{2} & 2^{-1/4} & \sqrt{3/\sqrt{2}-2} \\
2^{-1/4} & 1/2 & 1/2-1/\sqrt{2} \\
\sqrt{3/\sqrt{2}-2} & 1/2-1/\sqrt{2} & \sqrt{2}-1/2
\end{array}
\right).
\end{equation}
After passing through the network, the state of the three modes will be very complicated, but we then measure the number of photons contained in modes $1$ and $2$. If we measure the state $|1_1,0_2\rangle$, then the operation $NS_{-1}$ was successfully performed. The probability that we measure the desired result is $1/4$. If some other number of photons is detected then the operation has failed, and the information contained in mode $0$ has been destroyed.

One of the fundamental universal gates for quantum computation is the controlled sign flip, which shifts the phase of a two qubit state when both qubits have the logical value of $1$ but does nothing if one of the qubits is $0$. In our LOQC notation, this is the transformation
\begin{eqnarray}
|0_1,1_2,0_3,1_4\rangle & \rightarrow & |0_1,1_2,0_3,1_4\rangle \\
|0_1,1_2,1_3,0_4\rangle & \rightarrow & |0_1,1_2,1_3,0_4\rangle \\
|1_1,0_2,0_3,1_4\rangle & \rightarrow & |1_1,0_2,0_3,1_4\rangle \\
|1_1,0_2,1_3,0_4\rangle & \rightarrow & -|1_1,0_2,1_3,0_4\rangle.
\end{eqnarray}
To execute this we must first prepare four ancilla modes containing the state $|1_5,0_6,1_7,0_8\rangle$. Then we follow the procedure:
\begin{enumerate}
\item Apply the beam splitter ${\bf B}_{\pi/4,0}$ to modes $1$ and $3$.
\item Perform the $NS_{-1}$ operation on modes $1$, $5$, and $6$.
\item Perform the $NS_{-1}$ operation on modes $3$, $7$ and $8$.
\item Apply the beam splitter ${\bf B}_{-\pi/4,0}$ to modes $1$ and $3$.
\item Measure the number of photons in modes $5$ through $8$.
\end{enumerate}
If the state $|1_5,0_6,1_7,0_8\rangle$ is detected, then the controlled phase shift has been performed correctly. If some other state is detected, then the qubits have been destroyed. This will succeed with probability $(1/4)^2=1/16$, so we call this protocol for the controlled sign flip ${\rm c}-z_{1/16}$. Soon we will discuss other protocols that will increase the success probability, but these are more easily understood in the context of quantum teleportation.

The goal of quantum teleportation in a LOQC is to transfer the photon state of one mode to some other mode without directly connecting the two through a beam splitter. The following procedure will transfer the state $|\psi\rangle=a|0_0\rangle+b|1_0\rangle$ of mode $0$ to mode $2$.
\begin{enumerate}
\item Prepare the state $|1_1,0_2\rangle$.
\item Apply the beam splitter ${\bf B}_{\pi/4,0}$ to modes $1$ and $2$. This creates the state $|t_1\rangle=1/\sqrt{2}(|0_1,1_2\rangle+|1_1,0_2\rangle)$.
\item Apply the beam splitter ${\bf B}_{-\pi/4,0}$ to modes $0$ and $1$.
\item Measure the number of photons in modes $0$ and $1$.
\end{enumerate}
If the measurement yields $|1_0,0_1\rangle$ the mode $2$ will contain the state $a|0_0\rangle-b|1_0\rangle$, so to return this to the original state $|\psi\rangle$, apply the phase shifter ${\bf P}_{\pi}$ to mode $2$. If the measurement yields $|0_0,1_1\rangle$ mode $2$ will contain $|\psi\rangle$ and nothing must be done. If the measurement has some other result, the operation has failed. This procedure will succeed with probability $1/2$, but it can be increased with the following method:
\begin{enumerate}
\item Prepare the state
\begin{equation}
|t_n\rangle=\frac{1}{\sqrt{n+1}}\sum_{i=0}^{n}|1\rangle^{i}|0\rangle^{n-i}|0\rangle^{i}|1\rangle^{n-i},
\end{equation}
where $|x\rangle^i=|x\rangle|x\rangle|x\rangle...$ $i$ times. $|t_n\rangle$ requires $2n$ optical modes, and they should be numbered from left to right in each term of the sum. $|t_n\rangle$ can be prepared using only beam splitters and the state $|1\rangle^{n}|0\rangle^{n}$.
\item Operate on modes $0$ through $n$ with the beam splitter network ${\bf F_n}$, where the elements of the matrix describing ${\bf F_n}$ are $(F_n)_{j,k}=(e^{i2\pi/(n+1)})^{jk}/\sqrt{n+1}$
\item Measure the photons in modes $0$ through $n$.
\end{enumerate} 
If the number of photons detected is $m$, and $1\leq m \leq n$, then the teleportation procedure has succeeded. The original state occupying mode $0$ is found in mode $n+m$, but this state may need to be corrected with a known phase shifter. The probability that $1\leq m \leq n$ is $n/(n+1)$.

To improve the success of the controlled sign flip, we will perform the teleportation procedure on each qubit, but rather than use two copies of the state $|t_n\rangle$ we will use the state
\begin{equation}
|{\rm cs}_n\rangle=\sum^n_{i,j=0}(-1)^{(n-i)(n-j)}|1\rangle^{i}|0\rangle^{n-i}|0\rangle^{i}|1\rangle^{n-i}|1\rangle^{j}|0\rangle^{n-j}|0\rangle^{j}|1\rangle^{n-j}.
\end{equation}
Knill, Laflamme, and Milburn give an algorithm for the preparation of this state in \cite{KLM1}. According to their algorithm, the preparation of $|{\rm cs}_n\rangle$ requires $6n-3$ controlled phase shift operations \cite{error} each of which succeeds only probabilistically by performing the ${\rm c}-z_{1/16}$ operation or by preparing $|{\rm cs}_m\rangle$ for $m<n$ and executing this teleportation method. They provide no proof that their method is optimal, and we believe that finding the optimal procedure for preparation of $|{\rm cs}_n\rangle$ is a fruitful open problem. Imagine that qubit one occupies modes $q_1$ and $q_2$, qubit two occupies modes $q_3$ and $q_4$, and $|{\rm cs}_n\rangle$ occupies modes $1$ through $4n$. The algorithm is:
\begin{enumerate}
\item Operate on modes $q_1$ and the first $n$ modes of $|{\rm cs}_n\rangle$ with the beam splitter network ${\bf F_n}$.
\item Operate on modes $q_3$ and the third $n$ modes of $|{\rm cs}_n\rangle$ with ${\bf F_n}$.
\item Measure modes $q_1$ and $1$ through $n$. If $1\leq k_1 \leq n$ photons are detected. The contents of mode $q_1$ are now found in mode $n+k_1$.
\item Measure modes $q_3$ and $2n+1$ through $3n$. If $1\leq k_2 \leq n$ photons are detected, then the contents of mode $q_3$ are now found in mode $3n+k_2$.
\item Depending on $k_1$ and $k_2$, apply correcting phase shifters to $n+k_1$ and $3n+k_2$.
\end{enumerate}
Provided that $k_1$ and $k_2$ satisfy the above conditions, this procedure will perform the controlled sign flip on the qubits rather than simply teleporting them.  The probability that both teleportations succeed is $(n/(n+1))^2$, and we will refer to this procedure as ${\rm c}-z_{(n/(n+1))^2}$.  To increase this probability beyond 95\% would require $n \geq 39$, and to prepare the necessary state $|{\rm cs}_{39}\rangle$ involves $231$ uses of the controlled sign flip operation each of which succeeds probabilistically.

\section{Simulation of LOQC with imperfect detectors}
In order to investigate the effects of imperfect photon detectors on the operation of the LOQC, we have developed a computer simulation using Mathematica. \cite{Mathlink} The simulation represents the state of the LOQC in the Fock basis in which a basis state assigns an integer number of photons to each optical mode. Then superpositions are built out of the basis states in the usual manner.  The simulation can then calculate how the operation of a collection of beam splitters and phase shifters transform the states. To simulate ideal measurement the simulation will generate a table displaying the probability to measure all possible numbers of photons in each mode being measured, the number of measured photons, and the collapsed state produced after the measurement. At this stage the simulation has calculated the probabilities that the measured modes of the computer contain 0, 1, 2, ..., or $N$ photons, we must now calculate the probability that these photons are registered by the detectors.

We model photon loss using a binomial distribution. \cite{Mandel} If $N$ photons enter a detector, the probability that $n$ photons are registered by the detector is
\begin{equation}
p_{loss}(n,N,l)=\left(\begin{array}{c}N\\n\end{array}\right) \left(1-l\right)^n l^{N-n}
\end{equation}
when $n\leq N$, and $p_{loss}(n,N,l)=0$ when $n>N$. The inefficiency of the detectors is described with the parameter $l$ which is the probability that any given photon will be missed by the detector.

The photon detections may also be corrupted by noise photons.  To calculate the probability that $n$ noise photons are added to the number of ``real'' photons each detector registers, we apply the Poisson distribution.
\begin{equation}
p_{noise}(n,g)=\frac{g^ne^{-g}}{n!},
\end{equation}
where $g$ is mean number of noise photons each detector is expected to register.  Note that this Poisson distribution does not represent ``quantum noise'' in the photon field.  Instead, we imagine it to represent accidental, for example thermal, excitations inside the detector which are independent of the incoming photons.

The exact physics of the atom-photon interactions in a photon detector are very complicated, and we make no attempt to model them here. Our goal is only to describe the interactions using the two phenomenological parameters $l$ and $g$. It is difficult to imagine how a more detailed understanding of photon detector physics would yield wildly different probability distributions, so we are confident that our treatment will give a good understanding of the reliability of LOQC.

We can apply these probability distributions to the photons entering each detector to build a table that shows the probability that each detector registers 1, 2, 3, ... photons and the number of ``real'' photons that arrived at each detector. By adding the probabilities for the cases in which the number of photons registered by each detector and the number of photons that arrived at each detector are equal to one another and are both equal to the number prescribed by the LOQC protocol we can calculate $p_s(l,g)$ the probability that a given operation will succeed and the detectors will correctly detect the success. We also calculate $p_d(l,g)$, the probability that the detectors report that a given operation has succeeded, when in fact it may have failed, by adding the probabilities that the detectors register the number of photons prescribed by the LOQC protocol without regard to the number of ``real'' photons that arrived at the detectors.  Using $p_s$ and $p_d$, we can also calculate $p_f(l,g)=1-\frac{p_s}{p_d}$ the probability that the detectors register a false positive result, when the detectors register the number of photons prescribed by the LOQC protocol but the number of photons that arrived at the detectors is not equal to the prescribed number. This is surely the most damaging sort of error, because it is disguised as a successful calculation.

\begin{figure}[]
\centerline{\psfig{figure=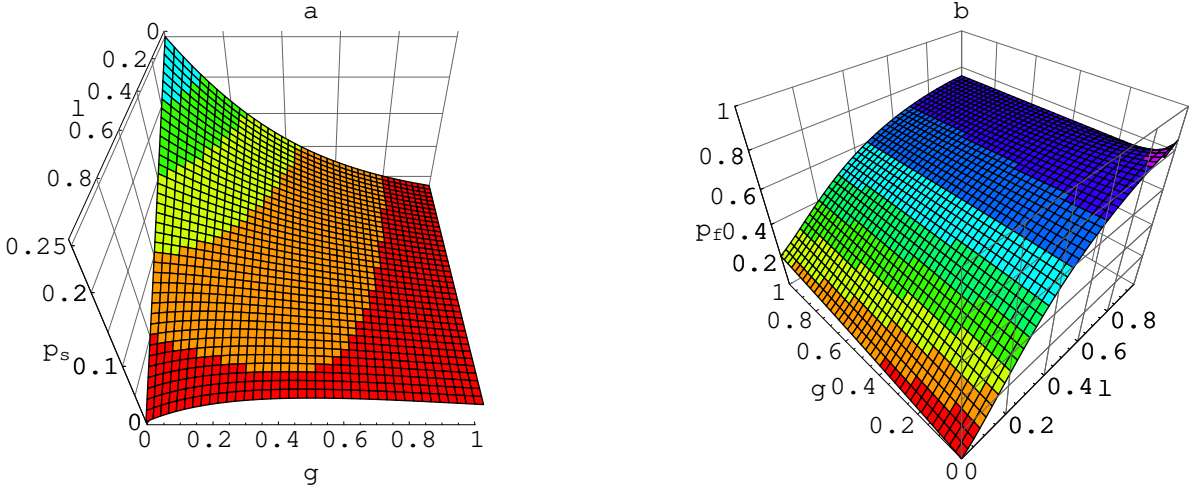}}
\caption{(a) shows the probability that the $NS_{-1}$ will succeed and the detectors will register the success plotted as a function of the detector inefficiency $l$ and the noise level $g$, for the case when the input mode contains two photons. (b) shows the probability that $NS_{-1}$ has failed, but the detectors register a success.}
\label{ns}
\end{figure}

We first applied this analysis to the $NS_{-1}$ operation, which succeeds when one photon is detected in mode $1$ and zero photons are detected in mode $2$. For an ideal device this should happen with probability $\frac{1}{4}$. The results of our simulation appear in Fig.\ \ref{ns}, where we have plotted the $p_s(l,g)$ and $p_f(l,g)$ for the case when the input mode contains two photons.

\begin{figure}[]
\centerline{\psfig{figure=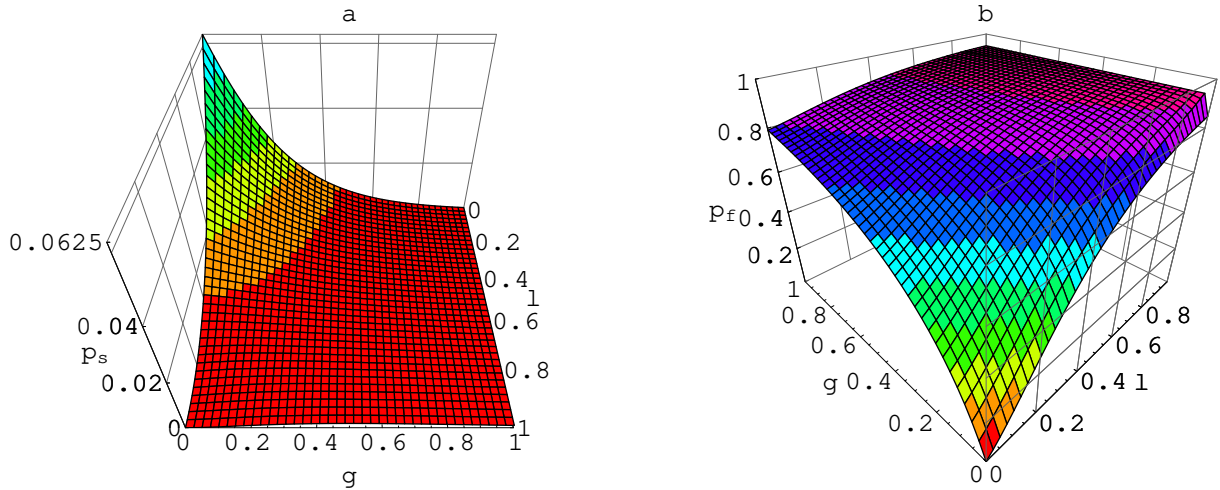}}
\caption{(a) shows the probability that the ${\rm c}-z_{1/16}$ will succeed and the detectors will register the success plotted as a function of the detector inefficiency $l$ and the noise level $g$, for the case when both input qubits are in the $|1\rangle$ state. (b) shows the probability that ${\rm c}-z_{1/16}$ has failed, but the detectors register a success.}
\label{cz116}
\end{figure}

We have also applied a similar simulation to the conditional sign flip ${\rm c}-z_{1/16}$, when both entering qubits are in the logical $|1\rangle$ state. Plots showing the success probability and false success probability are shown in Fig.\ \ref{cz116}. We find that for detectors with $l=0.1$ and $g=0.1$, the operation will succeed in only 3.5\% of its operations, and 35\% of its apparent successes are in fact failures. The probability of obtaining false positive results will be below 1\% for detectors better than those for which $l=g=0.0025$; $l=0$ and $g=0.0033$; or $l=0.0099$ and $g=0$.

\begin{figure}[]
\centerline{\psfig{figure=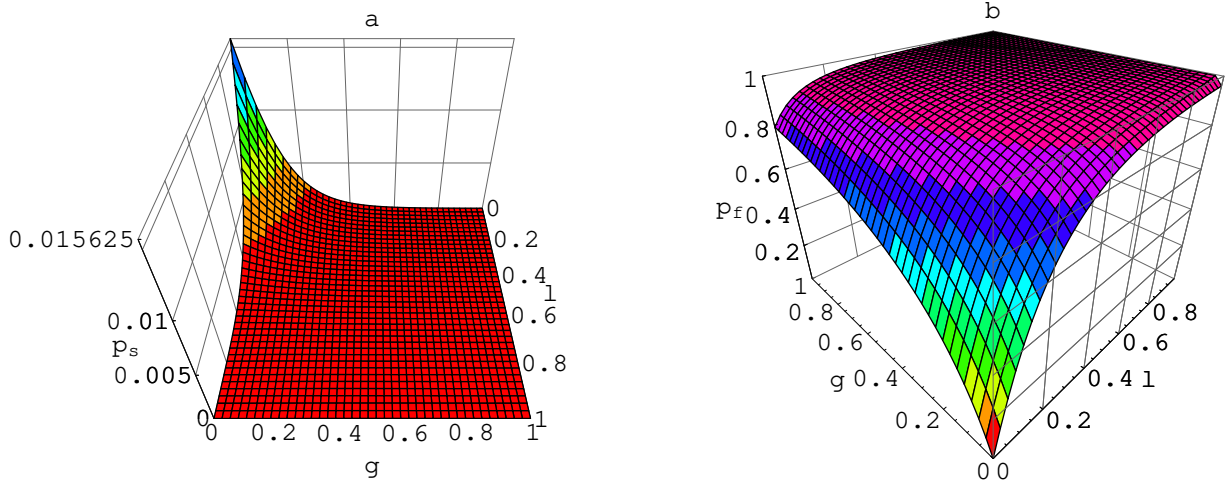}}
\caption{(a) shows the probability that the ${\rm c}-z_{1/4}$ will succeed and the detectors will register the success plotted as a function of the detector inefficiency $l$ and the noise level $g$, for the case when both input qubits are in the $|1\rangle$ state. (b) shows the probability that ${\rm c}-z_{1/4}$ has failed, but the detectors register a success.}
\label{cz14}
\end{figure}

The third system to which we apply our simulation is the controlled sign flip, ${\rm c}-z_{1/4}$, in which the state $|{\rm cs}_1\rangle=\frac{1}{2}\left(-|11\rangle+|10\rangle+|01\rangle+|00\rangle\right)$ is prepared using a single operation of the ${\rm c}-z_{1/16}$ gate. The two qubits, both in the logical $|1\rangle$ state, are then teleported using the $|{\rm cs}_1\rangle$ state.  Using perfect components, the probability of preparing $|{\rm cs}_1\rangle$ is $1/16$, and the probability of teleporting both qubits is $1/4$. Given detectors with $l=g=0.1$ we find that the probability to prepare $|{\rm cs}_1\rangle$ is 0.035, slightly more than half of the ideal probability, and 42\% of the apparent successes are failures. Using these detectors the probability to both prepare $|{\rm cs}_1\rangle$ and to teleport the qubits is 0.0048, only one third of the ideal success rate.  False positive results can be expected in 58\% of the attempts. False positive rates lower than 1\% can be achieved with detectors better than $l=g=0.0012$, $l=0.0020$ and $g=0$, or $l=0$ and $g=0.0033$. The less ambitious goal of lowering the false positive rate to 10\% would be possible with detectors better than  $l=g=0.013$, $l=0.021$ and $g=0$, or $l=0$ and $g=0.036$. These results are plotted in Fig.\ \ref{cz14}

\section{Near deterministic teleportation with imperfect detectors}
\begin{figure}[]
\centerline{\psfig{figure=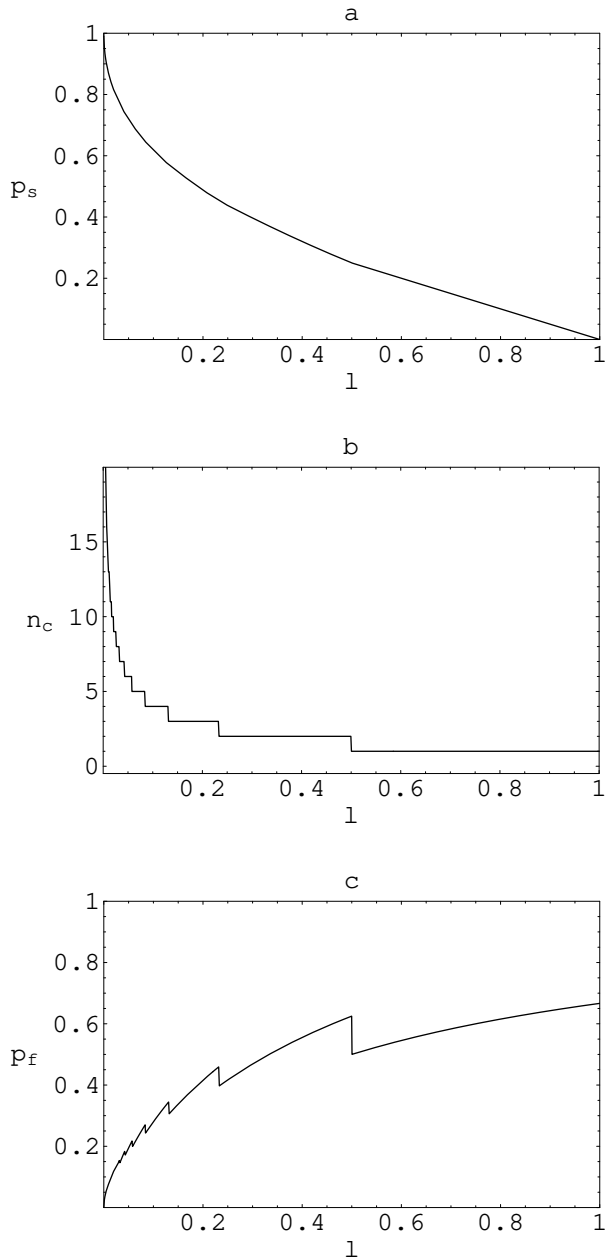}}
\caption{(a) shows the maximum probability that the teleportation will succeed and the detectors will register the success plotted as a function of the detector inefficiency $l$, for the case when the mode to be teleported contains one photon. (b) shows the value of $n_c$ that gives the maximum success probability. (c) shows the expected rate of false positive results when attempting a teleportation using a prepared state with size $n_c$.}
\label{teleport}
\end{figure}

Because quantum teleportation is such an integral part of the improved versions of the controlled sign flip ${\rm c}-z_{(n/(n+1))^2}$, we sought to understand the rate at which detector related errors grows as $n$ the size of the prepared teleportation state increases. Provided that the state $|t_n\rangle$ has been successfully prepared, we want to examine the probability that the state $|1_0\rangle$ can be teleported when using detectors with inefficiency $l$. For simplicity we assume that the detectors are noiseless.  By examining several cases we have found a general formula for $p_s(n,l)$ the probability that the teleportation succeeds and is correctly registered by the detectors:
\begin{equation}
p_s(n,l)=\frac{1}{n+1}\sum_{i=1}^{n}(1-l)^i=\frac{\left(1-\left(1-l\right)^n\right)\left(1-l\right)}{\left(1+n\right)l}.
\end{equation}
The probability that exactly $i=1, 2, ...,$ or $n$ photons arrive at the detectors is $\frac{1}{n+1}$, and the probability that all of the $i$ arriving photons are registered by the detectors is $(1-l)^i$. According to this expression, for a given detector inefficiency $l$, there is a critical $n=n_c$ for which $p_s$ is maximized, and increasing $n$ past $n_c$ will actually decrease the probability that the teleportation will succeed and be correctly registered by the detectors.  In Fig. \ref{teleport} we plot the maximum achievable $p_s$ and the $n_c$ yielding this probability for a range of detector inefficiencies.

\begin{figure}[]
\centerline{\psfig{figure=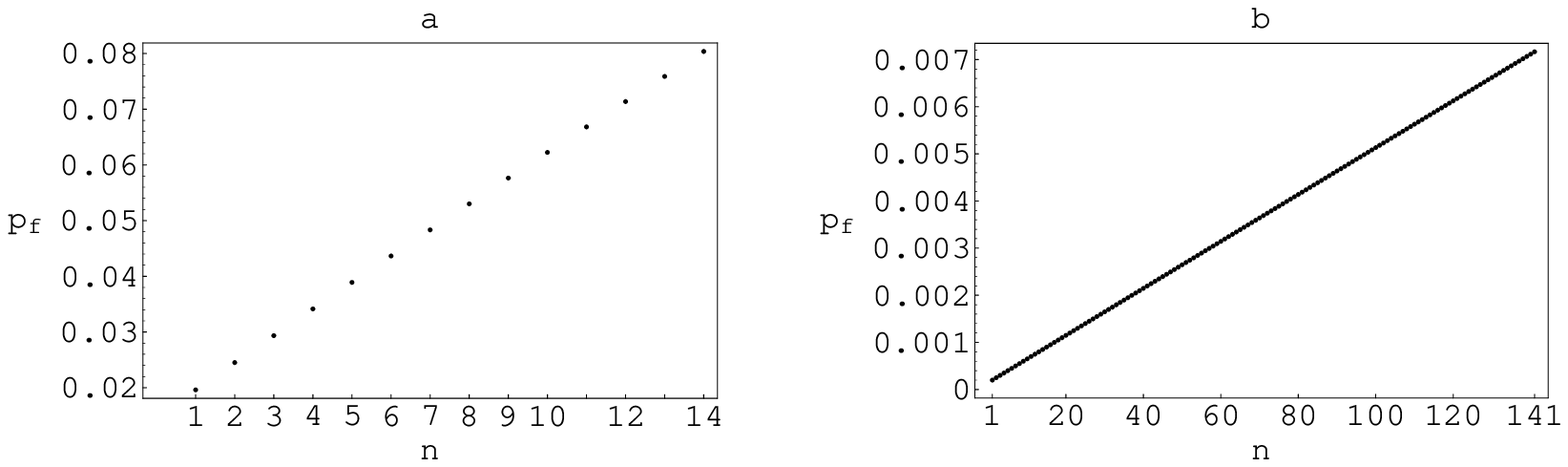}}
\caption{This shows the expected rate of false positive results as a function of the size of the prepared teleportation state for a device using detectors with inefficiency of 1\% in (a) and 0.01\% in (b).  The horizontal axes terminate at $n_c$, the size of the prepared teleportation state giving the maximum success probability for each detector.}
\label{teleport2}
\end{figure}

Of course, any increase in the number of detectors involved in the teleportation effort will increase the probability $p_f(n,l)$ that the detectors report that the teleportation has succeeded, when it has not.  To find $p_f(n,l)$ we first calculate $p_d(n,l)$ the probability that the detectors report that the teleportation was successful. This includes all of the events counted in $p_s(n,l)$ and all of the ways that photons may be lost so that the detectors register $1, 2, 3, ...,$ or $n$ photons.
\begin{equation}
p_d(n,l)=\frac{1}{n+1}\sum_{i=1}^{n}\left(\sum_{j=0}^{n+1-i}\left(\begin{array}{c}j+i\\j\end{array}\right)p^j\right)\left(1-p\right)^i
\end{equation}
$p_f(n,l)$ is then given by
\begin{equation}
p_f(n,l)=1-\frac{p_s(n,l)}{p_d(n,l)}.
\end{equation}
Any increase in $n$ will always increase the probability of receiving false positive results from the detectors. In Fig.\ \ref{teleport}c we have plotted $p_f(n_c,l)$. To build a teleportation device with a success probability of 0.9 requires detectors with $l=0.0055$ and $n=19$, and for 0.99 success rate requires $l=0.00006$ and $n=182$. Fig.\ \ref{teleport2} shows how quickly $p_f$ increases with $n$ for detector inefficiencies of 1\% and 0.01\%.

Construction of the near deterministic controlled sign flip requires the preparation of the state $|{\rm cs}_n\rangle$, which contains $2n$ photons distributed throughout $4n$ modes.  The first $2n$ modes are used to teleport the first qubit and the second $2n$ modes to teleport the second qubit, and after the teleportation the qubits will have received a controlled sign flip. The probability that each of these teleportations succeeds is equal to $p_s(n,l)$, and the probability that detectors report that each teleportation succeeds is just $p_d(n,l)$.  The probabilities of detecting the various possible numbers of photons during the teleportation of each qubit are independent of one another, so the probability to successfully implement the controlled sign is $\left(p_s(n,l)\right)^2$, and the probability of obtaining false positive results is
\begin{equation}
p_f(n,l)=1-\left(\frac{p_s(n,l)}{p_d(n,l)}\right)^2
\end{equation}
For detectors with $l=0.1$ the maximum achievable success rate is 0.38, which is obtained when $n=4$, but 48\% of apparent successes are false. To build a controlled sign teleportation that succeeds 90\% of attempts requires detectors with loss $l\leq 0.0014$ and $n\geq 37$.  A 99\% reliable gate needs $l\leq 0.000013$ and $n\geq 392$.

All of the results in this section rely on the assumption that the detectors produce zero noise. We chose to omit noise because of the complexity it adds to the calculations, because based on our simulations the errors caused by noise are very similar to those caused by loss when both the noise and loss are small, and because of the difficulty of displaying the results when varying inefficiency, $n$, and noise. Knill, Laflamme, and Milburn mention in \cite{KLM1} that one possible way to reduce the occurrence of false positive results in the teleportation procedure is to, after measuring $k$ photons in modes $0$ through $n$, also measure the photons in modes $n+1$ through $n+k-1$ and modes $n+k+1$ through $2n$ (all of the ancillary modes that do not contain the teleported qubit). Then the total number of detected photons should equal the number of photons in $|t_n\rangle$, otherwise one or more has been lost. Although this would reduce false positive results it will also reduce the device's success probability because sometimes the $n-1$ detectors in the second stage will lose photons when the $n+1$ primary detectors monitoring modes $0$ through $n$ function correctly, in which case good events would be discarded. If we naively imagine that each detector has the same probability $p$ to lose at least one photon, then using this method would reduce the success rate of the teleportation from $(1-p)^{n+1}$ to $(1-p)^{2n}$. From all successful teleportations, a fraction of $1-(1-p)^{n-1}$ would be rejected because some of the secondary detectors may malfunction.

\section{Conclusions}
In \cite{KLM2} Knill, Laflamme, and Milburn conservatively estimate that an accuracy threshold better than 99\% for controlled sign flip operations is required to construct a reliable LOQC. The main result of this paper is the calculation that to achieve this threshold when performing the ${\rm c}-z_{(n/(n+1))^2}$ with a perfectly prepared $|{\rm cs}_n\rangle$ state requires photon detectors with inefficiencies lower than $l=0.000013$, which are much superior to today's best photon detectors. This leads us to search for LOQC schemes that do not rely so heavily on the detection of single photons. One such possibility may be the use of coherent state qubits as outlined in \cite{Jeong}. In their scheme all detectors except one receive a large number of photons, and success or failure decisions are based on identifying the detector that registered zero photons compared to those detectors that registered a number of photons proportional to the (assumed to be large) amplitude of the coherent state.

This work has made no attempt to analyze the effect of unreliable photon sources and beam splitters, which would greatly complicate the analysis. Another open problem is the analysis of the many error correction procedures outlined in \cite{KLM1,KLM2,KLM3}, to understand their vulnerability to photon loss and noise. 

Lastly, we should note that this work should not be interpreted as a criticism of the many current experimental efforts \cite{Pittman} to build some basic LOQC elements.  Most of these experiments apparently plan to use coincidence measurements, which are largely impervious to corruption by photon loss or noise because those events can be discarded. However this is not a practical strategy to incorporate in a large scale quantum computer, because the qubits cannot be measured until the entire calculation involving very large numbers of controlled sign flip operations is complete.

\section{Acknowledgments}
S.~G. thanks the Arthur J.~Schmitt Foundation for fellowship support. J.~M.~L. is supported in part by the Division of High Energy Physics, Office of Science, U.~S.~Department of Energy under contract number DE-FG02-00ER41145. C. E. T. is supported in part by the Division of Chemical Sciences, Office of Basic Energy Sciences, Office of Energy Research, U. S. Department of Energy under contract number DE-FG02-95ER14579 and in part by the Atomic and Molecular Physics Program of the National Science Foundation under contract number PHY99-87984.


\begin{references}
\bibitem{error} Knill, Laflamme, and Milburn actually prescribe $6n-1$ instances of the controlled sign flip operations in \cite{KLM1}, but we have checked the algorithm and believe that they have included two which are unnecessary.\\
\end{references}
\end{document}